\documentclass[allclo]{FBSart}
\usepackage{amsfonts}
\usepackage{amssymb}
\usepackage{epsfig}

\title{Correlations in hot and dense quark matter} 

\author{S.  Mattiello$^a$\thanks{{\it E-mail address:}
    stefano.mattiello@physik.uni-rostock.de}, M. Beyer$^a$\thanks{{\it E-mail address:}
    michael.beyer@physik.uni-rostock.de}, T.
  Frederico$^b$, H. J. Weber$^c$}

\institute{$^a$Fachbereich Physik, Universit\"at Rostock, D-18051
  Rostock, Germany\\$^b$Dep. de F\'\i sica, Instituto Tecnol\'ogico de
  Aeron\'autica, Centro T\'ecnico Aeroespacial, 12.228-900 S\~ao Jos\'e dos Campos, S\~ao Paulo, Brazil\\
  $^c$Dept. of Physics, University of Virginia, Charlottesville, VA
  22904, U.S.A.}

\runningauthor{S. Mattiello}
\runningtitle{Few-quark correlation in hot matters}

\newcommand{\gl}{\lambda}

\begin{document}
\maketitle

\begin{abstract}
  We present a relativistic three-body equation to investigate
  three-quark clusters in hot and dense quark matter.  To derive such
  an equation we use the Dyson equation approach. The equation
  systematically includes the Pauli blocking factors as well as the self
  energy corrections of quarks.  Special relativity is realized
  through the light front form.  Presently we use a zero-range force
  and investigate the Mott transition.
\end{abstract}

%\vspace{5mm}   
%PACS: 12.39.Ki, %Relativistic quark model
%21.65.+f,       %Nuclear matter
%21.45.+v        %Few-body systems

%\vspace{5mm}
%\noindent
%Keywords: correlations, three-body equations, light front, quark matter,
%Mott transition, relativistic quark models, Dyson equations
\vspace{5mm}

\section{Introduction}

\begin{figure}[t]
\begin{minipage}{0.68\textwidth}
\epsfig{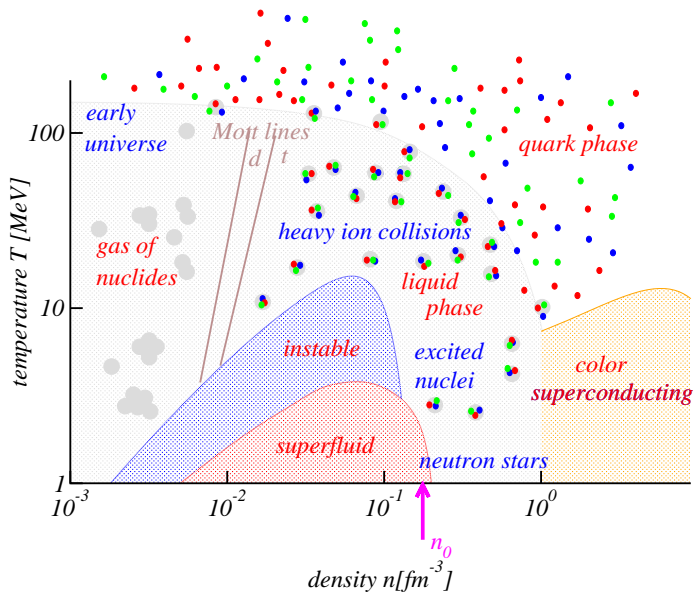}
\end{minipage}
\hfill
\begin{minipage}{0.3\textwidth}
\caption{\label{fig:phase} Sketch of the phase diagram of Quantum Chromodynamics.}
\end{minipage}
\end{figure}

There is increasing interest to investigate nuclear and quark
matter under a variety of conditions.  Results of lattice simulations
suggest a new state of matter, known as quark gluon plasma (see, e.g.,
Refs.~\cite{Karsch:Tc}). Recently, this
predicted state of matter has been claimed to have been observed by
experiments, see e.g. Ref.~\cite{Heinz:2000bk}.  Whereas
lattice calculations are largely progressing for zero chemical
potential ($\mu=0$) including dynamical quarks, see
e.g.~\cite{Ejiri:2001bw}, calculations at finite $\mu$ are more
difficult to implement, see e.g.~\cite{Ejiri:2001xg} for treatment with
a small finite chemical potential. In view of the present status of
lattice calculations it seems inevitable to employ further modeling
of QCD if one wants to explore the full phase diagram.  Effective
theories and models at various stages suggest a rich phase
structure at different temperatures and densities, see
e.g.~\cite{Alford:2001}. A sketch of the QCD density-temperature plane
is shown in Figure~\ref{fig:phase}.  Another interesting feature is
quark pairing that leads to the possibility of color superconductivity
analogous Cooper or nucleon pairing. The phase structure of QCD and in
turn the equation of state can be explored using specific signals in
heavy ion collision, see e.g. summary in~\cite{Blaizot:2001tf}. An
understanding of the QCD phases is also important for various
astrophysical questions, in particular to explore the structure of
neutron stars.

An aspect that has little been investigated in this context is the
appearance of three-quark correlations. We argue that in the vicinity
of the phase transition three-quark correlations should play an
important role since, below the critical temperature, nucleons (that
can be dominantly described as a bound state of three valence quarks)
are the relevant degrees of freedom. Three quark correlations might as
well influence quark pairing, a question that has recently been addressed
for the case $T=0$~\cite{Pepin}.

For a systematic investigation of three-quark correlations one needs
to derive a suitable relativistic three-body equation that includes
aspects of a medium at {\em finite temperatures and densities}. In the
past years utilizing the Dyson approach we have derived suitable
in-medium equations to treat three- and the four-correlations in a
medium of finite temperatures and densities, see
e.g.~\cite{Beyer:2001iz}. These equations are generalized here to
include special relativity.  To this end we use the light front
approach~\cite{Dirac:49}. Except for details in the treatment of the
medium the use of the light
front at finite densities is similar to that suggested in
Ref.~\cite{miller}.

\section{Theory}
For the time being we consider a zero-range interaction. A simple
effective theory of zero range is provided by the Nambu Jona-Lasinio
model~\cite{NJL}. Despite well known shortcomings, such as
missing confinement, this model reflects basic features of QCD. An
overview on the use of the NJL model in the context of finite
temperatures and densities can be found e.g. in Ref.~\cite{Klevansky}.
Note that due to screening effects the effective potential between
quarks may loose the confining property above the critical
temperature~\cite{Petreczky:2001pd}. Although the role of confinement
needs certainly to be clarified the main focus of our approach is on
the structure of the relativistic in-medium equation.

\subsection{Self energy-correction}
In Hartree-Fock approximation the self-energy correction induced by
the medium that changes the constituent mass $m$ is given by the gap
equation.  For the NJL model the gap equation leads to effective
masses $m(\mu,T)$ that depend on the temperature $T$ and the chemical
potential $\mu$ of the medium. For $T=10$ MeV, e.g. $m(100,10)=300$,
$m(200,10)=300$, $m(300,10)=289$, $m(307,10)=279$ (units
[MeV])~\cite{Schmidt:dipl,Mattiello:2001}.  As the chemical potential
increases for a given temperature the quark mass becomes smaller; same
for a given chemical potential while increasing temperature. To
proceed we use the light front formalism. As a Hamiltonian
formulation, this approach allows for a relativistic description of
the medium in terms of the statistical operator. The Fermi-Dirac
distribution function
\begin{equation}\label{f-Fermi} f(k^+,\vec
k_\perp)=\left(\exp\left[\frac{1}{k_BT}\left(\frac{k_{\rm on}^-+k^+}{2}
      -\mu\right)\right]+1\right)^{-1}  
\end{equation}
is expressed in terms of light front form momenta that are defined by
$\vec k_\perp=(k_x,k_y)$ and $k^\pm=k_0\pm k_z$. This treatment
coincides with the definition of blocking factors in the limit
$T\rightarrow 0$ given in Ref.~\cite{miller}. Note that the on
$k^-$-shell light front energy $k_{\rm on}^-=(\vec
k_\perp^2+m(\mu,T)^2)/k^+$ depends on $\mu$ and $T$.

\subsection{Two-body case}
The technical difficulties including angular momentum in relativistic
many-body systems are well known. For the time being we average over
the spin projections which means that the spin degrees of freedom are
washed out in the medium. This will be improved while the
investigation proceeds along the lines suggested in
Ref.~\cite{BKW:98}. Also we expect antiparticle degrees of freedom to
be of minor importance for a zero range interaction on the light
front~\cite{Frederico:1992}. In essence this leads to bose-type
relativistic few-body equations including medium effects. The solution
for the two-body propagator $\tau(M_2)$ for a zero-range interaction
is given by~\cite{Frederico:1992}
\begin{equation}
\tau(M_2)=\left(i\gl^{-1} - B(M_2)\right)^{-1}.
\label{eqn:prop2}
\end{equation}
where the expression for $B(M_2)$ in the rest system of the two-body system
is 
\begin{equation}
B(M_2)=-\frac{i}{(2\pi)^3} \int \frac{dx d^2k_\perp}{x(1-x)}
\frac{1-f(x,\vec k^2_{\perp})-f(1-x,\vec k_\perp^2)}{M_2^2-M_{20}^2} ,
\end{equation}
where $M_{20}^2=(\vec k_\perp^2+m^2)/x(1-x)$ and $x=k^+/P^+_2$.

\subsection{Three-body case}
The solution for the two-body propagator $\tau(M_2)$ is the input for
the relativistic three-body equation. The inclusion of finite temperature and
chemical potential is determined by the Dyson equations as explained,
e.g. in Ref.~\cite{Beyer:2001iz}. With the introduction of the vertex
function $\Gamma$ the equation becomes
\begin{eqnarray}
\lefteqn{\Gamma(y,\vec q_\perp) = \frac{i}{(2\pi)^3}\ \tau(M_2)
\int_{M^2/M_3^2}^{1-y} \frac{dx}{x(1-y-x)}}\nonumber\\
&&\int^{k_\perp^{\mathrm{max}}} d^2k_\perp
\frac{1-f(x,\vec k^2_\perp)
-f(1-x-y,(\vec k + \vec q)^2_\perp)}
{M^2_3 -M_{03}^2}\;\Gamma(x,\vec k_\perp) ,
\label{eqn:AGS}
\end{eqnarray}
where $m=m(\mu,T)$,
\begin{equation}
k_\perp^{\mathrm{max}}=\sqrt{(1-x)(xM_3^2-m^2)},
\end{equation}
and the mass of the virtual three-particle state in the rest system is
given by
\begin{equation}
M_{03}^2=\frac{\vec k^2_\perp+m^2}{x}
+\frac{\vec q^2_\perp+m^2}{y}
+\frac{(\vec k+\vec q)^2_\perp+m^2}{1-x-y} .
\end{equation}
The blocking factors, $1-f-f$ that appear in (\ref{eqn:AGS}) can be
rewritten as $\bar f\bar f -ff$, where $\bar f=1-f$ to exhibit the
particle and the hole blocking.  For $T\rightarrow 0$, $\bar f$
coincides with $\theta(k-k_F)$ that cuts the integrals below the
Fermi momentum.  Note that a blocking factor $\bar f\bar f\bar f$ as
suggested e.g. in~\cite{Pepin} leads to double blocking of the
spectator quark for a {\em two}-body kernel at {\em finite}
temperatures.

\section{Results}
\begin{figure}[b]
\begin{minipage}{0.48\textwidth}
\epsfig{figure=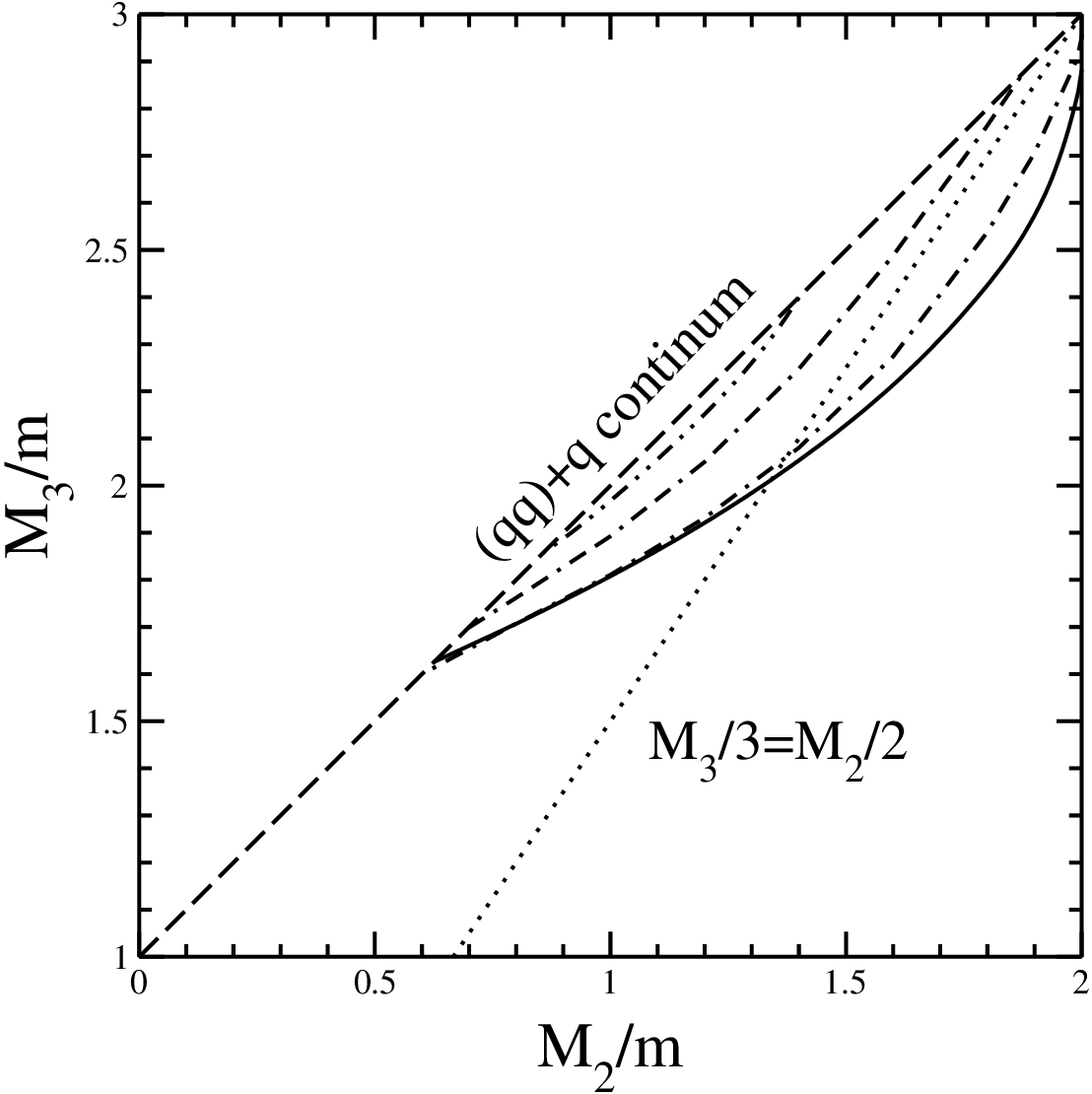,width=\textwidth}
\caption{\label{fig:M2M3} Masses of two-quark and
  three-quark bound states in units of quark mass at $T=10$ MeV for
  different chemical potentials $\mu$ (dashed-dotted line coding as in
  Figure~\ref{fig:medT10}). Further explanations see text.}
\end{minipage}
\hfill
\begin{minipage}{0.48\textwidth}
\epsfig{figure=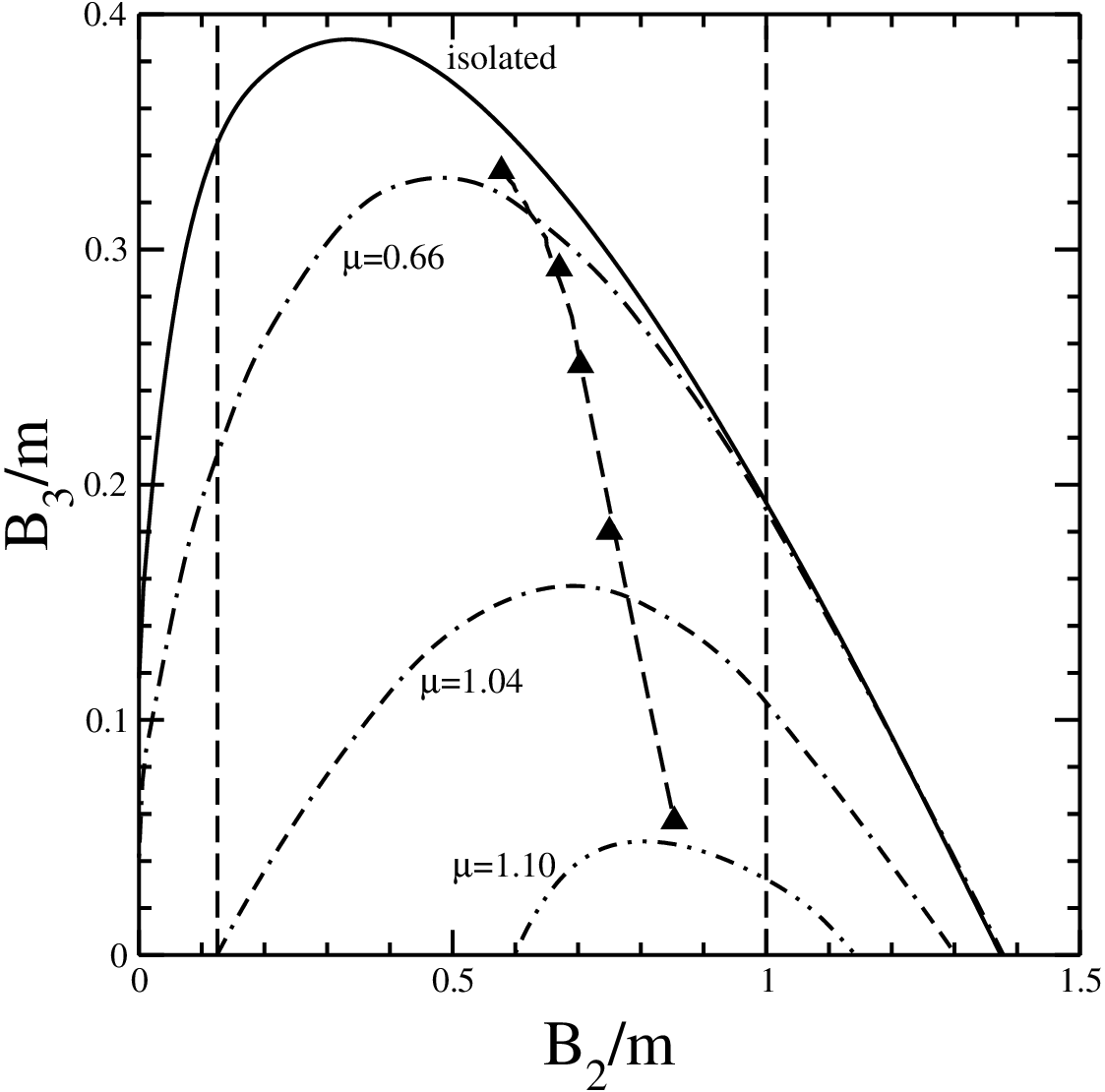,width=\textwidth}
\caption{\label{fig:medT10} Correlations between two-quark and
  three-quark binding energies in units of quark mass at $T=10$ MeV
  for different chemical potentials $\mu$, as indicated. Dashed with
  triangles \cite{Pepin}, further explanations see text.}
\end{minipage}
\end{figure}

For the time being we assume a bound state $M_{2B}$ in the two-body
subsystem.  This can be relaxed while the investigation proceeds and
more realistic models are implemented. Our main focus is on effects
the medium has on the competition between two and three-quark states.
This is particularly relevant in the vicinity of the critical
temperatures. A first step towards this aim here is the investigation of
the Mott transition from the three-body bound state to the
two body (2+1) channel. In passing we note that for the three nucleon
system the Mott transition of $^3$He or $^3$H is to the three-body
channel~\cite{Beyer:1999zx}. To this end, we vary $M_{2B}$ implicitly
choosing a particular model strength $\gl$ in (\ref{eqn:prop2}) for
each value of $M_{2B}$. This is shown in Figure~\ref{fig:M2M3}. For a
given $M_{2B}$ the solid line reflects the corresponding three-body
bound state for the isolated system. For a temperature of $T=10$~MeV
the various dashed-dotted lines correspond to increasing chemical
potential (see Figure~\ref{fig:medT10} for the corresponding values of
$\mu$). The long dashed line is the two-body continuum.  In a simple
chemical picture of an equilibrium system the dotted line indicates
the relative importance for the clusters due to the law of mass
\begin{figure}[t]
\begin{minipage}{0.48\textwidth}
\epsfig{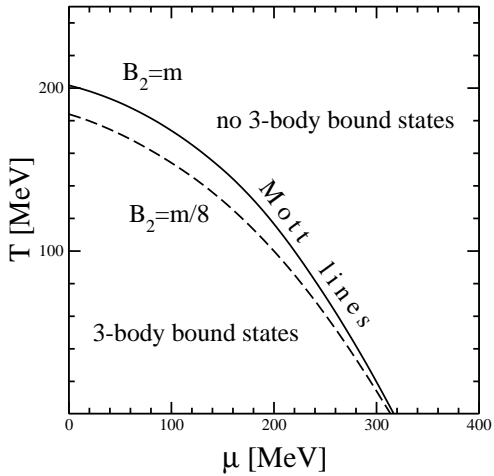}
\end{minipage}
\hfill
\begin{minipage}{0.48\textwidth}
\caption{\label{fig:mott}Mott lines for the three-body system at rest
  in the medium. For values of $T$ and $\mu$ below the Mott lines
  three-body bound states can be formed. Solid line for $M_2=m$,
  dashed line for $M_2=15m/8$.}
\end{minipage}
\end{figure}
action. In Figure~\ref{fig:medT10} we show the binding energies
\begin{eqnarray*}
  B_3(\mu,T)&=&m(\mu,T)+M_{2B}(\mu,T)-M_{3B}(\mu,T)\\
  B_2(\mu,T)&=&2m(\mu,T)-M_{2B}(\mu,T).
\end{eqnarray*}
The solid line refers to the isolated case and the various
dashed-dotted lines again represent the in-medium results for
different chemical potentials at $T=10$ MeV.  In addition, we refer to
the NJL results (dashed line with triangles) given in
Ref.~\cite{Pepin} for $m=450$ MeV and $T=0$: The triangles are for
values of $\mu(T=0)\simeq 1.0,1.05,1.08,1.12,1.22$ (top to bottom) in
units of the respective $m(\mu,0)$.  The dashed vertical lines reflect
specific values for $M_{2B}=m$ and $M_{2B}\simeq 1.88m$ in our model
covering a wide range of $M_{2B}$. For a given two-body binding energy
and increasing chemical potential we find weaker three-body binding.
This leads to the disappearance of the three-quark bound state for a
certain value of the chemical potential which is known as Mott
transition, $B_3(\mu_{\rm Mott},T_{\rm Mott})=0$. The values of $T$
and $\mu$ for which this transition occurs is plotted in
Fig.~\ref{fig:mott} for the two different models given above. Clearly
the behavior qualitatively reflects the confinement deconfinement
phase transition.

\section{Conclusion}

In conclusion, we have given a consistent equation for a relativistic
three-quark system in a medium of finite temperature and density. This
equation includes the dominant effects of the medium, viz. Pauli
blocking and self energy corrections. As many details need to be
improved, this first calculation shows that for a large range of
models the Mott transition agrees qualitatively with the phase
transition expected from other sources. Based on this approach it is
now possible to systematically investigate the influence of
three-quark correlations on the critical temperature and the onset of
color superconductivity at high density.

\begin{acknowledge}
Work supported by Deutsche Forschungsgemeinschaft.
\end{acknowledge}

\end{document}